# Adaptive Kinetic-Fluid Solvers for Heterogeneous Computing Architectures

Sergey Zabelok[a], Robert Arslanbekov[b], Vladimir Kolobov[b]

[a]*Dorodnicyn Computing Center of Russian Academy of Sciences, Vavilova-40, Moscow, 119333, Russia*
[b]*CFD Research Corporation, Huntsville, AL 35805, USA*
*vik@cfdrc.com*

**Abstract.** We show feasibility and benefits of porting an adaptive multi-scale kinetic-fluid code to CPU-GPU systems. Challenges are due to the irregular data access for adaptive Cartesian mesh, vast difference of computational cost between kinetic and fluid cells, and desire to evenly load all CPUs and GPUs. Our Unified Flow Solver (UFS) combines Adaptive Mesh Refinement (AMR) with automatic cell-by-cell selection of kinetic or fluid solvers based on continuum breakdown criteria. Using GPUs enables hybrid simulations of mixed rarefied-continuum flows with a million of Boltzmann cells with 24x24x24 velocity mesh. We describe the implementation of CUDA kernels for three modules in UFS: the direct Boltzmann solver using discrete velocity method, the Direct Simulation Monte Carlo (DSMC) solver, and a mesoscopic solver based on Lattice Boltzmann Method, all using octree Cartesian mesh. Double digit speedups on single GPU and good scaling for multi-GPUs have been demonstrated.

**Keywords:** Unified Flow Solver, Adaptive Mesh Refinement, Discrete Velocity Method, Boltzmann Kinetic Equation, Direct Simulation Monte Carlo, Lattice Boltzmann Method, Graphics Processing Units, CUDA, MPI.

## 1. INTRODUCTION

It is well known that transport phenomena can be described by either atomistic (kinetic) or continuum (fluid) models. Kinetic description in terms of particle distribution functions is more detailed and computationally more expensive compared to the continuum description in terms of density, mean velocity and temperature. Two methodologies have been used to solve the kinetic equations: statistical particle-based methods such as Direct Simulation Monte Carlo (DSMC) or Particle-in-Cell (PIC) [1,2] and direct numerical solutions using computational grid in phase space.[3,4] Continuum models are computationally efficient but have limited range of applicability. Mesoscopic models such as Lattice Boltzmann Method (LBM) attempt to bridge the gap between the two methods. Multi-scale kinetic-fluid models are being developed to enable using kinetic and fluid solvers in different parts of systems to achieve maximum fidelity and efficiency.[5,6,7,8,9,10] Adaptive kinetic-fluid models apply different solvers in dynamically selected regions of physical or phase space for efficient description of multi-scale phenomena in complex systems. Appropriate solvers are selected using sensors locally detecting phase space regions where kinetic approach is required and apply fluid models in other parts of the system.

For gas dynamics in mixed rarefied-continuum regimes, the adaptive kinetic-fluid approach has been first realized using an Adaptive Mesh *and* Algorithm Refinement (AMAR) methodology introduced for DSMC-fluid coupling.[11] Later, a Unified Flow Solver (UFS) has been developed to combine Adaptive Mesh Refinement (AMR) with automatic cell-by-cell selection of direct Boltzmann solver or Euler-Navier-Stokes solvers[12] based on continuum breakdown criteria.[5] The AMAR methodology has been extended for hybrid modeling of radiation transport,[13] and is now being developed for plasma simulations.[6]

Figure 1 shows the basic architecture of UFS. The AMAR core is implemented on top of Gerris Flow Solver (GFS) - an open source computing environment for solving partial differential equations with AMR.[14] GFS provides with automatic mesh generation for complex geometries, portable parallel support using the MPI library, dynamic load balancing, and parallel offline visualization. GFS physics includes time-dependent incompressible variable-density Euler, Stokes or Navier-Stokes equations with volume of fluid method for interfacial flows. A coarse-grained parallelization algorithm is based on "Forest of Trees" methodology.[15]

UFS enables the higher degree of adaptation by using different physical models in different parts of the computational domain. The computational domain is decomposed into kinetic and fluid cells using physics-based continuum breakdown criteria. This methodology was first implemented for mixed rarefied-continuum flows and later extended for hybrid model of radiation transport coupling a Photon Monte Carlo (PMC) solver with a diffusion model of radiation transport selected based on the local photon mean free path.[13]

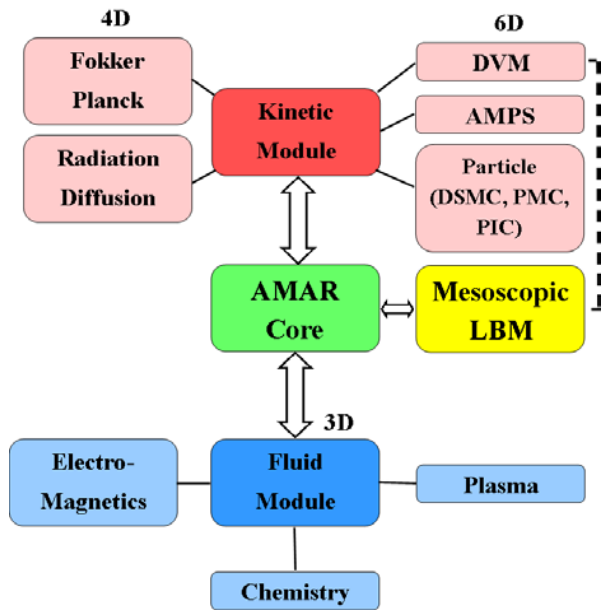

*Figure 1: The AMAR framework*

Kinetic Module in UFS can solve Boltzmann, Vlasov, and Fokker-Planck kinetic equations using different methods. Eulerian solvers use Discrete Velocity Method (DVM) for solving kinetic equations. The recently developed Adaptive Mesh in Phase Space (AMPS) methodology [16] can adapt mesh in both physical and velocity spaces. The DSMC, PMC and PIC modules are based on Lagrangian transport models.[17] The mesoscopic LBM solver uses a minimal set of discrete velocities as a subset of the DVM kinetic solvers.[18]

Fluid Module in UFS contains multi-species Euler and Navier-Stokes solvers for reacting gas mixtures based on the Roe approximate Riemann solver, an exact Riemann solver with a Godunov-type scheme, AUSMPW+ scheme with MUSCL reconstruction, and the gas-kinetic schemes.[12] For plasma simulations, multi-temperature drift-diffusion models for electrons and ions coupled to Poisson solver for the electrostatic field are used.[6]

This paper demonstrates feasibility and benefits of adapting the UFS framework for heterogeneous computing architectures using Graphical Processing Units (GPUs). GPUs have become powerful data-parallel computation accelerators for a wide range of devices from high-end supercomputers to battery-powered mobile devices, tablets, and laptops.[19] For programs that map well to GPU hardware, GPUs offer a substantial advantage over multicore CPUs in terms of performance, performance per dollar, performance per transistor, and energy efficiency.

The tree-based adaptive Cartesian mesh of UFS poses specific challenges for porting to GPU systems due to irregular data access. Furthermore, different cost of kinetic and fluid cells must be taken into account when developing parallel algorithms. Because of the high computational cost of the kinetic solvers and their suitability for GPU acceleration, we first implemented GPU algorithms for the kinetic solvers in UFS. In Section 2, we briefly describe parallel algorithms of the AMAR technique and our strategy for porting adaptive kinetic-fluid solvers to heterogeneous CPU-GPU systems. Section 3, describes GPU algorithms for the Boltzmann solver, shows example of GPU acceleration and scaling. At the end of Section 3, we show an example of hybrid kinetic-fluid simulations obtained on a cluster with multiple CPU-GPU nodes. Section 4 describes the implementation of CUDA kernels for the LBM solver. Although the LBM Module in UFS is a subset of the DVM kinetic solvers, the CUDA algorithm for LBM module is quite different compared to that of the Boltzmann solver. Section 5 describes DSMC solver and its porting to multi-GPU systems. We conclude with a summary of our results.

## 2. PARALLEL IMPLEMENTATION OF AMAR

The main idea of parallel algorithms implemented in UFS for both multi-CPU and multi-GPU computations is based on decomposition of computational domain in physical space into subdomains each corresponding to one device (CPU, GPU, or both). There are two ways to obtain efficient domain splitting and to have a good load balancing. The fine grained parallelization uses Space Filling Curves (SFC).[20] The coarse grained parallelization is based on Forest of Trees (FOT).[15] Both these approaches were first implemented and tested on multi-CPU clusters. In the present paper, these algorithms are applied for heterogeneous computers having both multiple CPU and GPU units. In the following subsections these algorithms are considered.

### 2.1 Fine Grained Parallelization using Space Filling Curves

The computational grid in UFS is generated by subsequent division of square boxes into smaller boxes with linear dimensions equal to half of the initial dimension (left part of Figure 2). The procedure of creating the computational mesh can be represented by a tree (the right part of Figure 2). The root of the tree ($0^{th}$ level) corresponds to initial cube, the first level corresponds to 4 (in 2D) or 8 (in 3D) cubes obtained by division of the initial cube, etc. The computational cells correspond to leaves of the tree. Two additional constraints are imposed to simplify the gradient and flux calculations: 1) the levels of direct neighbors cannot differ by more than one, and 2) the levels of diagonal neighbors cannot differ by more than one.

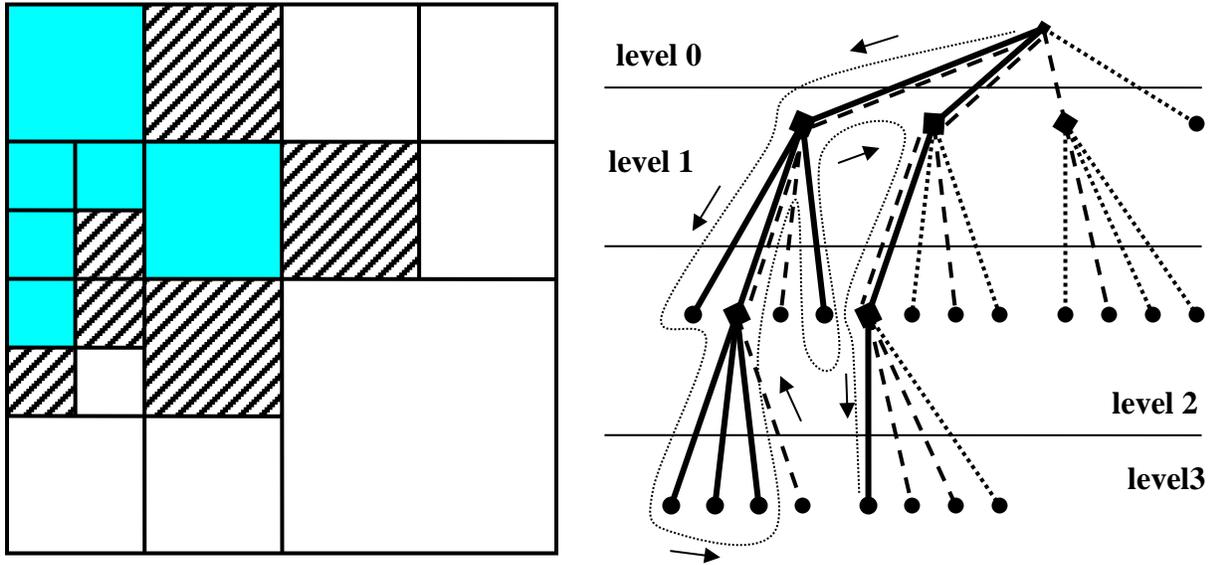

*Figure 2: The computational mesh and the tree corresponding to it.*

All computation procedures are called only for the leaves of the tree. To perform computations for a part of the domain (for instance, the sub-domain shown by blue color in Figure 2), one introduces a flag for each leaf to identify whether or not the cell belongs to the selected sub-domain. If a cell is a parent cell for a set of cells, it is flagged only if at least one child cell belongs to the selected sub-domain. After introducing flags, the procedure of cell traversing is modified to visit only the cells belonging to selected sub-domain. As shown on the right part of Figure 2, only the branches connected by solid lines are traversed. To specify boundary conditions, additional cells (hatched in Figure 2) are used. The branches corresponding to the boundary cells are shown by dashed lines on the graph. The boundary cells are marked using a different flag, and can be traversed separately by the code.

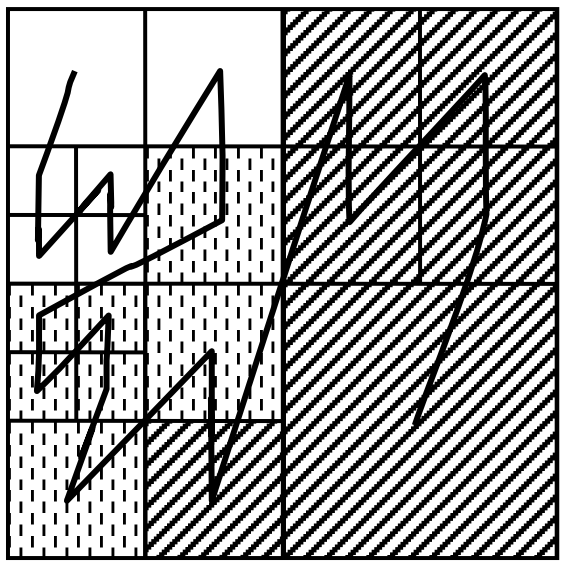

*Figure 3: Domain partitioning obtained with SFC algorithm for the case of 3 processors.*

Using SFCs, domain decomposition is dynamically performed as illustrated in Figure 3. During sequential traversing of all cells in a natural order, the physical space is filled with curves. Different types of SFCs can be used for this purpose. In UFS, we currently use N-order (Morton ordering). After ordering of the cells, a weight is assigned to each cell based on the node time required to perform computations in this cell. By node in this section we mean a MPI node with CPUs/GPUs attached to it. Kinetic cells have a much larger weight compared to the fluid cells. Furthermore, the array modified with corresponding weights, is subdivided into sub-arrays equal to the number of nodes, in such a way that the weights of each sub-arrays are approximately the same. This method allows very precise and efficient domain decomposition between different nodes and is called fine-grained parallel algorithm.

## 2.2 Coarse Grained Parallelization using Forest of Trees

The FOT algorithm consists of dividing the computational domain into boxes of fixed size with a Cartesian tree built in each box. Each tree becomes the smallest parallel subdomain. The trees are connected through their common boundaries. Graph partitioning algorithms are used for domain decomposition and dynamic load balancing (DLB). This represents a coarse-grain parallelization algorithm since the CPU load balancing can be done only in terms of the building (root) boxes. Each node operates on a given number of boxes, and the computational grid is stored only on the node it belongs to. During grid adaptation these boxes are exchanged between the nodes for DLB. Each box has its own ID number and a set of boxes on each host node share the same Processor or node ID (PID) number. The box-based DLB has the benefit of efficient memory allocation because only the trees that belong to a node need to be kept in the node's memory. A detailed discussion of advantages and drawbacks of the coarse-grain and fine-grain parallelism can be found in Ref. [15].

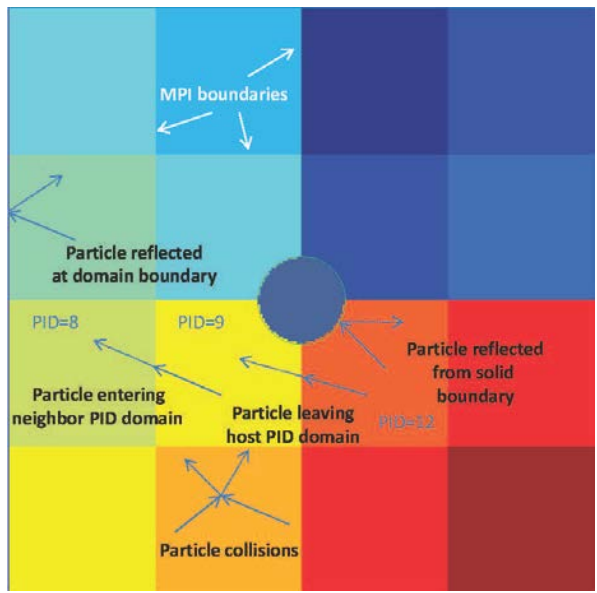

*Figure 4. PID partioning for 16 CPU/GPUs for the problem of heat transfer. Also shown are illustrations of particle interactions with solid surfaces, with other particles and with domain and MPI boundaries.*

The DSMC module in UFS uses the FOT parallelization algorithm. Figure 4 illustrates an example of initial domain decomposition into 16 boxes for DSMC simulation of rarefied gas confined between a cold cylinder inserted into a square box with hot walls. At the beginning of simulations, computational gas particles are initialized in each PID domain. These particles start to move and collide with each other, with domain boundaries, and with solid surfaces/phases immersed into the domain, as illustrated in Figure 4. Particles (their position, velocity, remaining move time, etc.) which hit an MPI boundary are stored in a special list and then removed all together from their PID domain. Knowing the direction of the box boundary face a particle hits, its neighbor PID (e.g., PID=8 in Figure 4) is determined. This neighbor PID is then used for the thus stored particles to send to and to receive from (PID=9 in Figure 4). The neighboring PIDs receive and then add these particles for further processing during the next time step. Then this procedure repeats for the next time step and so on.

### 2.3 Challenges of porting adaptive kinetic-fluid codes to CPU-GPU systems

Efforts required for porting existing computational algorithms and software to CPU-GPU architectures depends of the problem type. GPUs can be very effective at exploiting parallelism in regular programs that operate on large arrays or matrices and access them in statically predictable ways. These programs often have high computational demands, exhibit extensive data parallelism, and access memory in a streaming fashion requiring little synchronization. Such "low-hanging fruit" *regular* codes, which require no significant change in program structure, have already been ported to GPUs.[21] Irregular codes associated with dynamic data structures (such as graphs, trees, etc.) are more difficult to port. Nevertheless, GPUs are capable of accelerating many irregular codes, and there is plenty of exploitable parallelism available even in irregular codes.[22]

Many groups have worked to port CFD solvers into GPUs.[23] Expected acceleration depends on the type of fluid model and the grid type. Traditional CFD solvers are based on Navier-Stokes equations. Alternative methods use LBM and gas-kinetic schemes. They utilize elements of the kinetic theory to provide solutions beyond the Navier-Stokes equations and thus often called mesoscopic methods. The LBM is both computationally expensive and memory demanding, but its explicit nature and the data locality (the computations for a single grid node require only the values of the neighboring nodes) make it ideal for parallel implementations. Thus, LMB methods are best suited for GPU computing and are increasingly being used for simulations of low speed flows, especially for flows over complex and moving boundaries (cars, engines, porous media, etc.).

In terms of mesh technologies, adaptive Cartesian meshes offer attractive capabilities of automatic mesh generation for complex geometries with minimal user intervention and dynamics mesh adaptation not achievable with the traditional body-fitted mesh techniques. Moreover, AMR minimizes the number of computational cells for high resolution of local flow features such as shock waves, boundary and mixing layers, etc. On the other hand, they result in dynamic data structures (octrees, graphs) typical to irregular codes, which are more difficult to port on GPUs.

Among successful irregular CFD codes designed for GPUs is the open source code, GAMER[24] (GPU-accelerated Adaptive MEsh Refinement), which is based on a hierarchy of grid patches with an octree data structure. The GAMER framework has AMR, a variety of GPU-accelerated hydrodynamic and Poisson solvers, hybrid OpenMP-MPI-GPU parallelization, concurrent CPU-GPU execution for performance optimization, Hilbert SFC for load balance. However, GAMER does not have capabilities for inserting and treatment of static or moving solid objects into the computational domain, and uses a fixed number of cells ($8^3$) in each regular patch. As discussed in Ref. [24], restricting all patches to be geometrically similar to each other greatly simplifies the AMR framework with respect to GPU implementation. A single GPU kernel can be applied to all patches, even in different refinement levels. Moreover, since the amount of computation workload of each patch is the same, there is no synchronization overhead when multiple patches are evolved in parallel by GPU.

Kinetic solvers are well suited for parallel computing. Impressive acceleration and excellent multi-GPU scaling have been demonstrated by several groups using particle [25,26] and grid-based methods with structured meshes. [27,28] We have recently demonstrated double digit speedups on a single GPU [29] and good multi-GPU scaling for the DVM Boltzmann solver, the DSMC module, and the LBM solver, all using adaptive Cartesian meshes.[18] Coupled Vlasov and two-fluid code has been recently described using GPU for Vlasov solvers in locally selected kinetic domains.[9]

Porting adaptive multi-scale kinetic-fluid solvers such as UFS to heterogeneous CPU-GPU computing architectures poses several challenges associated with irregular data structures of the dynamically adapted mesh, vastly different costs of computing in fluid and kinetic cells, and dynamic balancing of both CPU and GPU loads. We use here the simplest approach where GPUs are used only for the kinetic cells whereas the fluid cells are computed on CPUs. This methodology is demonstrated for UFS simulations of mixed rarefied-continuum flow over a hypersonic scramjet on a CPU-GPU cluster. The compressible NS solver is used in the fluid cells, whereas the DVM Boltzmann solver is used in the kinetic sells. The cell-by-cell selection of the kinetic or fluid solvers is based on the continuum breakdown criteria using the local Kn number.

It would be possible to couple LBM and Boltzmann solvers in UFS for simulations of low speed flows using LBM in regions with low local Kn numbers and the DVM Boltzmann solver in the areas with larger Kn numbers. We have not tried this option so far, but do not anticipate obstacles using it in future work. Hybrid DSMC-NS simulations have been described in literature (see [30] and references therein). Such simulations appear particularly attractive for high-speed (hypersonic) flows with large variations of the flow macro-parameters and the velocity distribution function within computational domain. In fact, the AMAR acronym was first introduced in Ref [11] for hybrid DSMC-fluid simulations. However, coupling particle-based kinetic solvers with grid-based fluid solvers involves some challenges at the kinetic-fluid interfaces. We have not explored this option in UFS so far, but plan to use it in future work for plasma simulations.

## 3. BOLTZMANN SOLVER

The UFS-Boltzmann solver is based on the direct numerical solution of the Boltzmann kinetic equation (BKE). For one-component atomic gas, BKE has the following form (in non-dimensional variables):

$$\frac{\partial f}{\partial t} + (\vec{\xi}, \overrightarrow{\nabla_x} f) = \frac{1}{Kn} I(f,f). \qquad (1)$$

Here *f* is the 7 - dimensional velocity distribution function, which depends on time *t*, and on two 3-dimensional vectors (coordinate $\vec{x}$ and molecular velocity $\vec{\xi}$), and *I* is a collision integral with quadratic dependence on the velocity distribution function. The Knudsen number *Kn*, which is defined as the ratio of the mean free path to the characteristic spatial scale, is a measure of the gas rarefaction degree.

The numerical solution of BKE is based on the discrete velocity method (DVM). The distribution function is defined at cell centers of uniform Cartesian mesh in velocity space and adaptive non-uniform Cartesian mesh in physical space. The left-hand side of BKE (advection operator) is evaluated using finite volume technique; the right-hand side (collision integral) is computed using Korobov nodes in velocity space (see details in Ref. [5]). Other methods that can be used for computing collision integral are reviewed in Ref. [31].

This numerical scheme has been successfully ported to GPU and demonstrated good speedup on single GPUs.[29] For many problems, the computational times for the collision and advection parts of the Boltzmann solver were of the same order and it was necessary to develop both advection and collision CUDA kernels for GPU computing. In many other methods, implemented on GPU[32,33], the time for the collision integral computations was much larger than for the advection, so it was not necessary to construct an effective advection scheme.

The method for numerical solution of BKE with CUDA GPU is based on specific presentation of the required data. In it the distribution function points are reordered in such a way that they form vectors for which the same arithmetical operations are performed. This type of vector data presentation allows effective use of CUDA GPU capabilities.

### 3.1 Advection Algorithm

The algorithm for approximation the advection operator on GPU is based on the properties of FV advection scheme used in UFS for approximating convective part of BKE. This numerical scheme can be written as follows:

$$\frac{f_{i,j}^{n+1} - f_{i,j}^n}{\Delta t} + \xi_{x,i} \sum_{k=1,\ldots,m_x(i,j)} a_x(k,i,j) f_{i,\alpha_x(k,i,j)}^n + \xi_{y,i} \sum_{k=1,\ldots,m_y(i,j)} a_y(k,i,j) f_{i,\alpha_y(k,i,j)}^n +$$
$$\xi_{z,i} \sum_{k=1,\ldots,m_z(i,j)} a_z(k,i,j) f_{i,\alpha_z(k,i,j)}^n = 0. \qquad (2)$$

Here *n* corresponds to the time step, *i* and *j* enumerate the points of discrete mesh in velocity and physical space respectively. In the case of a Cartesian mesh the coefficients in the above formula (at a given point of physical space) are the same for molecular velocities belonging the same octant (the components of molecular velocities have the same signs), i.e. for the numbers $m_\beta(i,j)$, coefficients $a_\beta(k,i,j)$ and indexes $\alpha_\beta(k,i,j)$ $\beta=x, y, z$ the dependence on the index $i$ can be replaced with the dependence on $\text{sign}(\xi_{\beta,i})$. This is used in algorithm for numerical advection scheme.

The algorithm for CUDA computations consists of the following steps:

1. Create 8 **groups** of velocity points with the same signs of velocity components. The first group corresponds to signs (-, -, -), the second one – to (-, -, +), etc. Rearrange the points of velocity space in such a way that for a given point each **group** to be located in one consecutive block of memory.

2. Create an array for the distribution function storage.

3. Store the distribution function for different points of physical space in a single array, with the points in velocity space to be ordered in the same way as it is done in step 1.

4. For each point in physical space $j$ store numerical scheme for each **group** of velocity points, i.e. record the quantity of stencil points $m_\beta(sign\ (\xi_{\beta,i}),j)$, stencil point numbers $\alpha_\beta(k, sign(\xi_{\beta,i}),j)$, the stencil weights $a_\beta(k, sign(\xi_{\beta,i}),j)$.

5. Copy all the data obtained in steps 2 and 4 to device memory.

6. Execute the advection kernel.

7. Copy results from device to host memory.

Step 1 is done once before the 1-st time step. Steps 2 and 4 are redone each time the spatial mesh is changed. Other steps are executed every time step.

The advection kernel (step 6) takes the addresses of arrays of the distribution function and stencil parameters as parameters. A single thread block computes the advection contribution of all velocity points of one **group** of velocity points for one point of physical mesh, with each thread corresponding to a point of velocity space. The stencil parameters are copied to shared memory. The molecular velocity components, which are different for each thread, are computed once per kernel execution and are stored in the register memory. Thus the procedure of computing the advection operator consists of a sequence of vector multiplication (the distribution function for points of one velocity **group**) by a number from shared or register memory operations, which is a very effective operation for SIMD devices. For CUDA GPUs this procedure also allows to avoid problems with coalescing.

### 3.2 Collision Integral

The collision integral for BKE is approximated by the following formula:

$$I_{i,j}^{*n} = -f_{i,j}^n \sum_{k=1,\ldots,M(i)} b(k,i) f_{\gamma(k,i),j}^n + \sum_{k=1,\ldots,M(i)} c(k,i) f_{l(k,i),j}^n f_{m(k,i),j}^n. \tag{3}$$

The number of trials $M(i)$, the indexes $\gamma(k,i)$, $l(k,i)$, $m(k,i)$ and the coefficients $b(k,i)$, $c(k,i)$ are the same for all points of velocity space. The star superscript denotes that this formula gives an intermediate approximation of the collision integral, which does not satisfy the conservation laws. In order to satisfy the mass, momentum and energy conservation, the computations of the collision integral are divided into 3 steps:

1. Computations of the collision frequency
$$B_{i,j}^n = \sum_{k=1,\ldots,M(i)} b(k,i) f_{\gamma(k,i),j}^n \tag{4}$$

2. Computations of the inverse collisions integral
$$C_{i,j}^n = \sum_{k=1,\ldots,M(i)} c(k,i) f_{l(k,i),j}^n f_{m(k,i),j}^n. \tag{5}$$

3. Conservative correction
$$I_{i,j}^n = -f_{i,j}^n P_{i,j}^n B_{i,j}^n + C_{i,j}^n, \tag{6}$$

$$P_{i,j}^n = \left(1 + d_0(j) + d_x(j)\xi_x(j) + d_y(j)\xi_y(j) + d_z(j)\xi_z(j) + d_2(j)\vec{\xi}^2(j)\right).$$

The five coefficients $d_\beta(j)$ of the polynomial $P_{i,j}^n$ for each spatial point $j$ in the last operation (conservative correction) are found from the system of five linear equations, which are numerically equivalent to the conservation laws.

The algorithm for CUDA computations consists in the following steps:
1. Create an array for the distribution function storage.

2. Store the distribution function as an array of vectors with each vector corresponding to a single point of velocity space and to all points of physical space.

3. Compute the number of trials $M(i)$, the indexes $\gamma(k,i)$, $l(k,i)$, $m(k,i)$ and the coefficients $b(k,i)$, $c(k,i)$ and store them in a corresponding array.

4. Copy arrays from steps 2 and 3 to device memory.

5. Create arrays for device storage of collision frequency and inverse collision integral.

6. Execute kernel for computations of collision frequency and inverse collisions integral.

7. Reorder the distribution function, collision frequency and inverse collision integral in device memory, for them to become the arrays of vectors with each vector corresponding to a single point of physical space and to all points of velocity space. This is done by a simple kernel which perform transpose operation.

8. Execute conservative correction kernel.

9. Copy the results to host memory.

In the collision kernel (step 6) a single thread block computes the collision frequency and inverse collisions integral for the entire physical mesh for one point of velocity space, with each thread corresponding to a cell in physical space. The parameters $M(i)$, $\gamma(k,i)$, $l(k,i)$, $m(k,i)$ $b(k,i)$, $c(k,i)$ are the same for all threads inside the thread block, and thus they are copied to the shared memory. Thus procedure for computing collision frequency and inverse collisions integral consists of sequence of multiply vector (the distribution function for all points of physical mesh for a given velocity point) by a number from shared memory operations, which is (the same as for advection operator) very effective operation for SIMD devices, and also allows to avoid problems with coalescing for CUDA GPUs.

In the conservative correction kernel a thread block corresponds to a cell in physical space. At first the matrix elements and the right-hand side vector of the 5x5 linear system are computed inside each block. These elements are obtained as moments of the distribution function. Every thread inside the block computes the contribution to a moment of one velocity point, and after that the sum of all contributions is gathered to the thread with number 0 with a $O(log_2[number\_of\_threads])$ operations. The next step is to solve the 5x5 linear system which is easily done with the Gauss method using 1-st 25 threads of block. The last operation in the kernel is to obtain the value of collision integral. For this the vector of polynomial $P_{i,j}^n$ values is constructed (only shared and register memory is used for polynomial computations). After that the component multiplication and addition of vectors $P_{i,j}^n$, $f_{i,j}^n$, $B_{i,j}^n$ and $C_{i,j}^n$ is done.

Computational tests demonstrated speedup of advection kernel itself was from 15 to 40 times for different CPU and GPU combinations. The calculations of the collision integral in UFS-Boltzmann-GPU have demonstrated speedups of up to ~50. [18]

### 3.3 Multi-GPU Implementation of the Boltzmann Solver

A multi-GPU algorithm was implemented for UFS-Boltzmann solver for the particular hardware architecture with one GPU attached to one CPU. The SFC method of UFS was used to uniformly partition the computational domain (in physical space) between processors. Partitioning in physical space allowed leaving the collision integral computations unchanged, because the collision integral is calculated locally.

The advection algorithm includes the following items. The memory storage for the distribution function is arranged in such a way that the distribution function of the cells from the considered CPU (inner cells) are stored before the cells, which are included in a stencil for this CPU but belong to other CPU (ghost cells). The distribution function for ghost cells is gathered from other CPUs using MPI procedures. The device memory is arranged in the same manner as the host memory, the distribution function is copied from host to device using cudaMemcpy() function. The advection kernel, which was previously developed for a single GPU code, is called to process only the part of device memory corresponding to inner cells (but this kernel utilizes the ghost cells distribution function). The latest step of the algorithm is to gather the advection contribution from device to host. The scheme for this algorithm is presented in Figure 5.

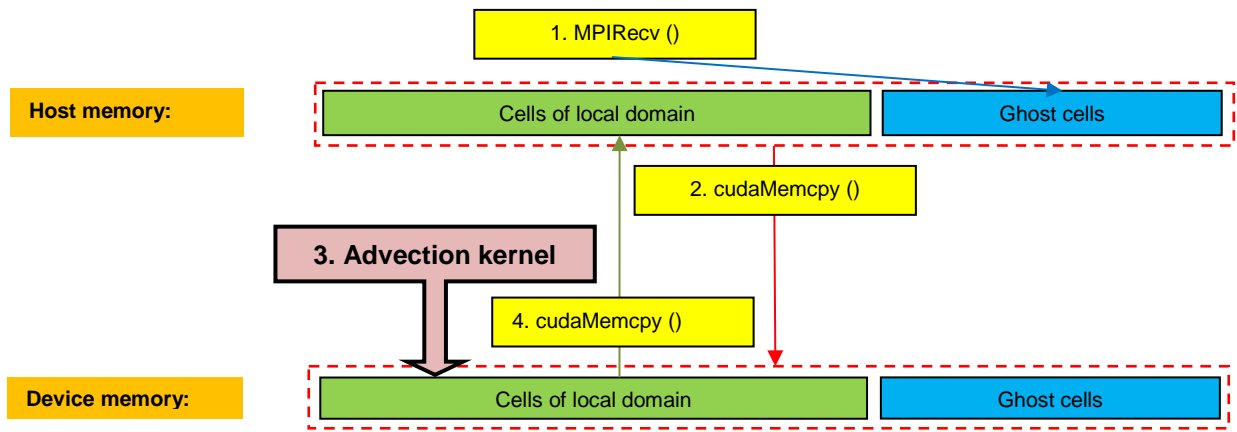

*Figure 5. Schematic algorithm for MPI-CUDA implementation of the advection part of the UFS-Boltzmann solver.*

The described algorithm was first tested on a two CPU-GPU node computer for a 3D problem of rarefied subsonic flow over a sphere at M = 0.1, Kn = 0.01. The speedup compared to a single node run was 1.8. Then, we simulated the same problem on the NASA Pleiades cluster with GPU M2090 nodes (older GPU cards) and GPU K40 nodes (the most recent cards*).* Figure 6 shows the GPU/CPU speedup obtained as function of the number of M2090 GPU cards for the velocity grid 32x32x32 with the number of spatial grid cells about 15K. For one GPU, speedup factors are 14 for advection and 83 for collision.

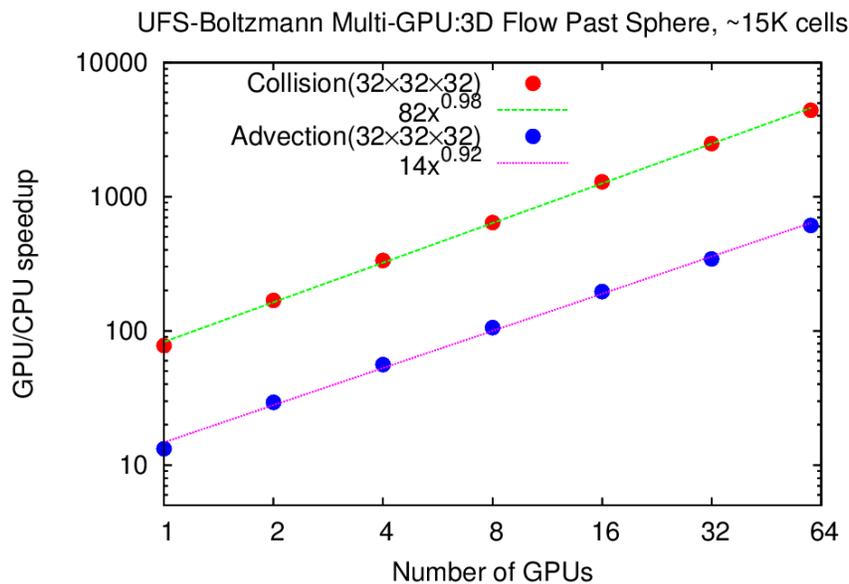

*Figure 6. GPU acceleration of the Boltzmann solver as function of the number of GPU cards for M2090 cards.*

Next, we simulated the same problem with the velocity grid 24×24×24 and the physical space grid of about 75K cells. Figure 7 compares the times required to execute the advection (transport) and collision operators for 1 time step. One can see that almost ideal speed up scaling is observed with the number of GPU devices on both M2090 and K40c cards. For the most powerful K40 cards, the collisional integral executes almost twice as fast, while the advection operator does not show acceleration. More detailed analysis of the timings shows that the time taken to prepare and then finalize data for the advection operator becomes larger than the advection kernel computing time. The kernel-only time does reduce by a factor of 2 when going from an M2090 card to a K40 card. We thus conclude that the recent GPU cards are so fast that the actual computing time inside the kernel becomes negligible compared to the time required for data preparation and transfer. This result calls for an all-on-device computation algorithm, which will be considered in future implementations of the UFS kinetic solvers.

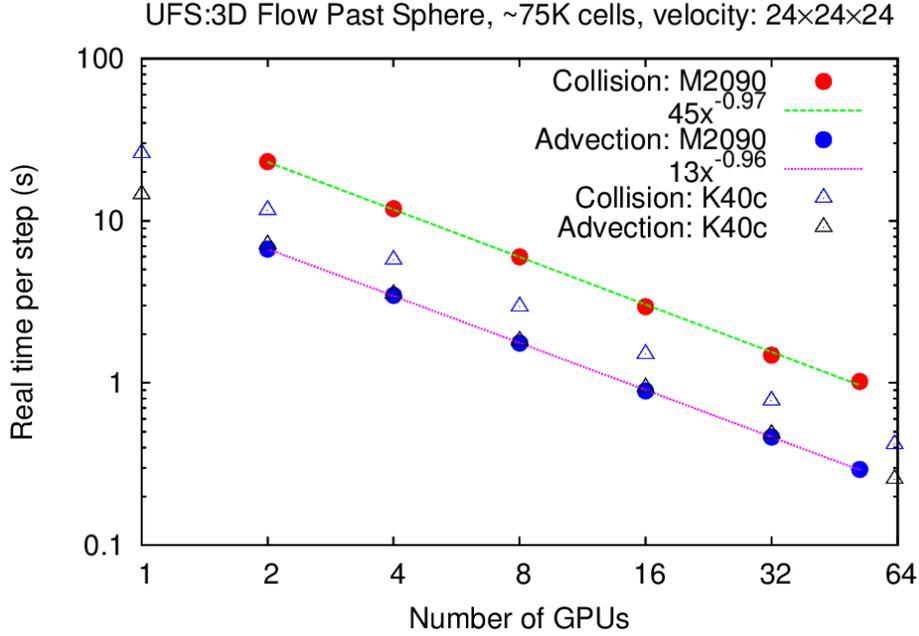

*Figure 7. The times per one step required for the collision and advection operators as a function of the number of GPU cards for two types of cards, M2090 and K40c.*

### 3.4 Example of hybrid CPU-GPU simulations of rarefied-continuum flow

Our algorithm for hybrid, multiscale kinetic-fluid simulations is based on CPU computations of the Navier-Stokes cells and GPU computations of the Boltzmann cells. This seems to be a reasonable approach due to high computational cost of the Boltzmann cells compared to the fluid cells. In this section, we illustrate this methodology for UFS simulations of one-component gas flow over a X-51A waverider scramjet at $Ma = 5$ and $Kn = 0.01$ (per vehicle length). The compressible NS solver was used in the fluid cells, whereas the DVM Boltzmann solver with hard sphere collision model was used in the kinetic sells. The cell-by-cell selection of the kinetic or fluid solvers was based on the continuum breakdown criteria using the local Kn number, which was defined through gradients of density, mean velocity and temperature as described in Ref [5]. Coupling kinetic and fluid solvers through the interface was performed assuming Maxwellian velocity distribution of the gas particles in the fluid cells at the kinetic-fluid interface. In turn, flow macro-parameters in the kinetic cells required for setting boundary conditions for the NS solvers were calculated from the VDF.

Figure 8 shows results of simulations performed on a NASA cluster using 58 computing nodes with GPU cards M2090. The total number of adapted computational cells is 3.5M, and the total number of kinetic cells (shown by pink color in Figure 8, right) is ~1.2M. The kinetic cells were activated in the regions of large local Knudsen number based on continuum-kinetic switch (see Ref. [5] for details). Static uniform velocity mesh 24x24x24 was used for the Boltzmann solver. Domain decomposition in physical space was obtained using SFC method. Using GPUs in the kinetic cells to speed up the Boltzmann solver's advection and collision modules we observed double digit speedup for the kinetic cells and good DLB between the nodes thanks to the SFC parallelization algorithm.

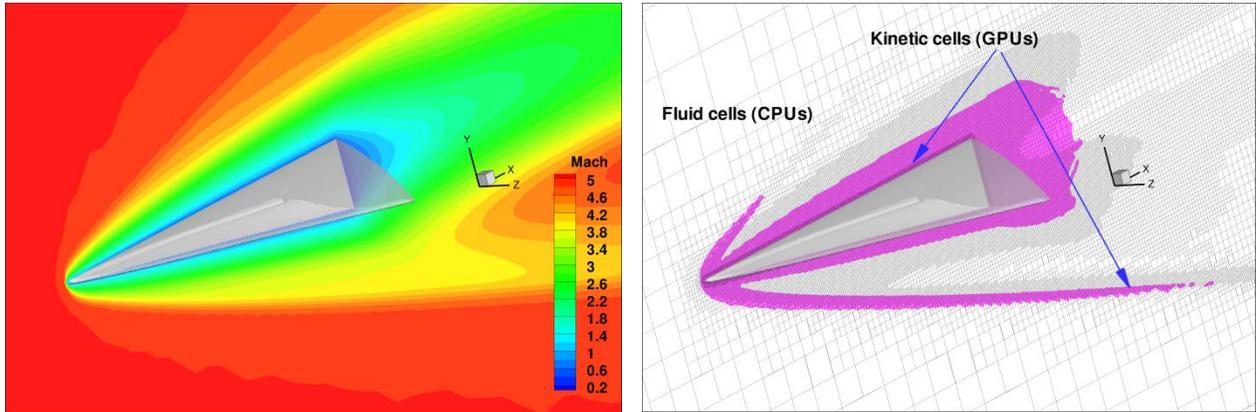

*Figure 8. Mach number (left), computational mesh and kinetic domain (brown cells, right) around a Waverider scramjet.*

This example demonstrated feasibility of porting adaptive multi-scale kinetic-fluid solvers to heterogeneous CPU-GPU systems. Using GPUs helped achieving a computational breakthrough for hybrid simulations of gas flows with total number of adapted computational cells about 3.5M, and total number of Boltzmann cells about ~1.2M. Solving Boltzmann equation with a million of spatial cells appeared unrealistic a few years ago.

The results described above, have been obtained for a one-component monatomic gas using coupled DVM-Boltzmann and NS solvers. For multi-component gas mixtures and weakly ionized gases (plasmas), it might be desirable to use both grid-based and particle-based kinetic solvers in UFS. Below, we describe the CUDA implementation of the LBM and DSMC solvers, which we plan to use in future for simulations of these systems.

To improve the performance of hybrid NS-Boltzmann solver described above one can execute one part on CPU and another on GPU simultaneously, e.g. this could be done using asynchronous calls of GPU kernel. This seems to be a good way because modern CPUs are rather powerful, and if a part of the code is not well ported to GPU it can be executed on CPU is there is a job for GPU at this time.

## 4. LATTICE BOLTZMANN METHOD

The LBM was originally designed as an alternative solver for computational fluid dynamics. Later, it has been extended beyond the level of the Navier Stokes hydrodynamics, and capable to describe some kinetic effects.[34] LBM operates with a Velocity Distribution Function (VDF) defined on a minimal set of discrete velocities to obtain governing equations for the fluid dynamics alternative to conservation equations based on VDF moments. Most LBM works are devoted to low speed isothermal flows close to equilibrium. Recently, larger sets of discrete velocities [35] and adaptive meshes in physical space,[36,37] have been introduced to expand LBM capabilities. A number of LBM solvers, some of them with GPU capabilities, have been developed, e.g. Palabos,[38] LB3D,[39] Sailfish,[40] HemeLB,[41] LUDWIG[42] and Musubi.[43] However, to the best of our knowledge, no LBM codes with both AMR and GPU capabilities have been reported so far.

The LBM Module in UFS is designed as a subset of the DVM kinetic solvers. The DVM kinetic solvers allow using velocity grids (lattices) of arbitrary size N×M×L. The LBM module uses 3×3 (D2Q9), 19 (D3Q19), and 3×3×3 (D3Q27) velocity lattices. The finite volume LBM implementation in UFS follows that of Ref. [44]. Special procedures and structures map UFS's regular Cartesian velocity grid to the LBM velocity grids. These structures also contain the weights of each LBM velocity vector required to compute the equilibrium distribution function for initial and boundary conditions and for the BGK collisional model.

The BGK collisional term has the standard discrete form for each velocity grid point $\vec{\xi}_i$ ($i = 0, \ldots, N$ with $N$ the lattice size):[44]

$$I_i^{BGK} = -\frac{1}{\tau}(f_i - f_i^{eq}),$$

where $\tau$ is the relaxation time and $f_i^{eq}$ is the local equilibrium function, which in the second order expansion can be written as

$$f_i^{eq} = w_i \rho \left(1 + \frac{\vec{\xi}_i \vec{u}}{c_s^2} + \frac{(\vec{\xi}_i \vec{u})^2}{2c_s^4} - \frac{u^2}{2c_s^2}\right),$$

where $w_i$ are the weights associated with the lattice velocities $\vec{c}_i$ and $c_s$ is the lattice speed of sound (=$1/\sqrt{3}$). The flow mass and momentum are computed as: $\rho = \sum_{i=0}^{N} f_i$ and $\rho\vec{u} = \sum_{i=0}^{N} \vec{c}_i f_i$, correspondingly. The BGK collisional operator is nonlinear because the quantities $f_i^{eq}$ are functions of local macroscopic flow variables $\rho$ and $\vec{u}$.

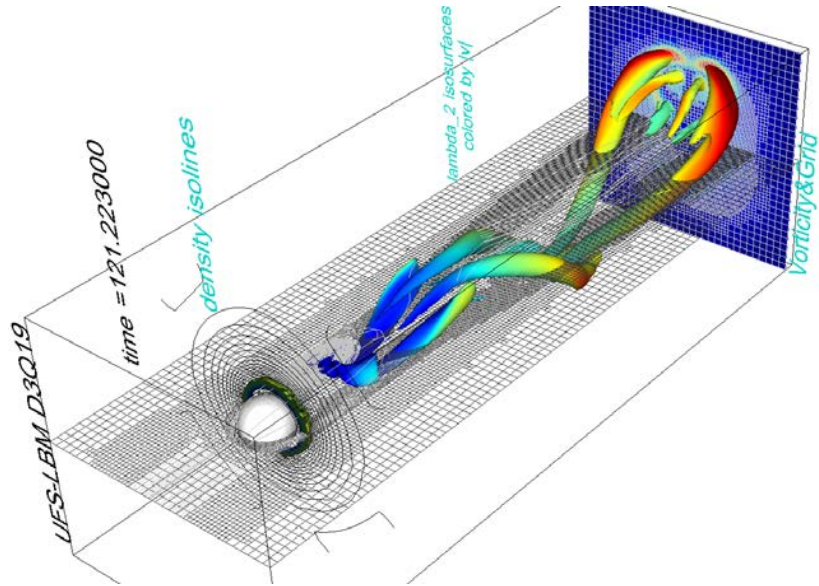

*Figure 9. Complex 3D structure of the periodic wake obtained with UFS-LBM D3Q19 model. The vertical cross-section is colored according to the vorticity. The horizontal and vertical cross-sections display the depth of refinement of the adaptive mesh.*

The SFC algorithm is used for parallelization of the LBM module. Figure 9 shows an example of parallel simulations of gas flow over a sphere at Ma = 0.5 and Re = 1,000 with D3Q19 model on eight processors. The obtained complex structure of the periodic wake is illustrated by iso-surfaces of the (eigenvalue) lambda2 criterion of Jeong and Hussain.[14] One can see that computational grid adapts well to the flow vorticity, and the DLB algorithm allows keeping approximately the same numbers of cells per processor, even when the number of total cells increases more than 10 fold during the run; a close to ideal processor load is observed (see Figure 10). The lambda2 criterion helps visualize the formation of hairpin vortices in 3D flow over a sphere shown in Figure 9. The features of the flow resemble closely those obtained for a similar problem with static, regular Cartesian grid by the MFIX code.[45]

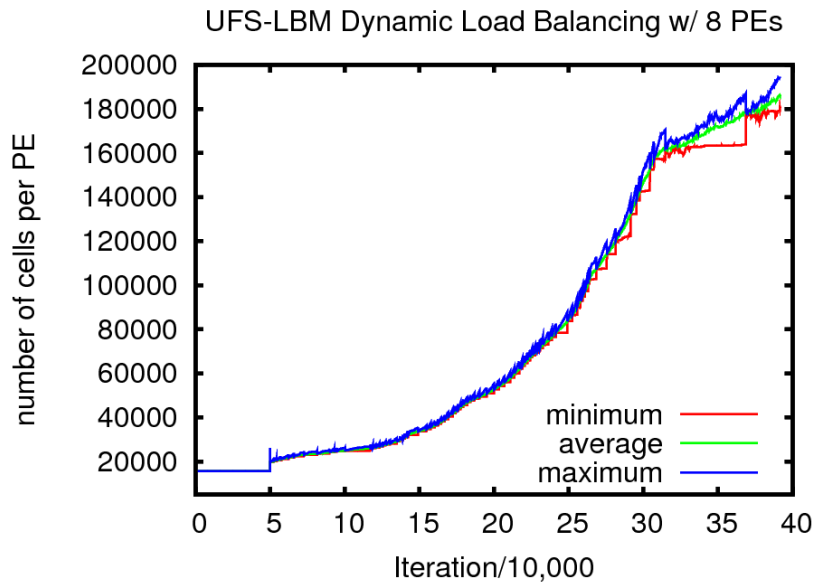

*Figure 10. Processor load (number of cells per processor) during UFS-LBM simulations of a 3D gas flow over a sphere on eight processors.*

### 4.1 GPU accelerated LBM Solver

GPU acceleration of the transport step in LBM is the most crucial because the collision step takes only 20% of the total computational cost. The main idea behind the LBM-tuned kernels developed in this work relies on the fact that the number of velocity grid points is small in LBM and the number of spatial cells can be very large. This allows us to group each velocity grid in a spatial cell into a warp. The concept of thread groups or warps is fundamental to the GPU's operation; warps are groups of threads handled in parallel by the GPU's execution units. For a D3Q27 model, the number of velocity grid points is 27 and the warp size is 32. This means that almost the whole warp will be processing the velocity grid points in parallel (with only 5 threads remaining idle). We then group 16 or 32 warps into a block and the number of blocks is decided upon the total number of spatial cells. Since the transport module in UFS-LBM involves summations over neighboring cells (which can be any number from 6 to 6×4 on an Octree mesh depending on the level of refinement of each particular spatial cell), a warp which finishes computations first will switch to another warp and so on. As pointed out in Ref. [46], when one warp stalls on a memory operation, the multiprocessor selects another ready warp and switches to that one; in this way, the cores can be productive as long as there is enough parallelism to keep them busy. This allows the computations to proceed without delay on a large number of cells of different neighbor topologies, which is particularly important for an octree mesh.

GPU speedup has been tested for the same 3D flow past a sphere presented above. We have verified that the obtained CPU and GPU results coincide within a machine precision. We compared times required to execute one transport and one collisional step on CPU and on GPU for two types of grids with ~100K and ~1M cells. Table 1 shows results for two GPU cards (Fermi and Kepler) (results are for kernel-only computations discarding pre- and post-processing times). One can see that the advection kernel shows speedup factors ~40 for the Tesla C2070 and ~60 for the Tesla K20c card. As expected, the speedup factors are larger for the Tesla K20c because it is a more powerful GPU. The collisional kernel achieves even larger speedup factors of ~150 for both cards. However, the collisional kernel takes a small fraction of the total advection-collision step, and the overall contribution of the collisional kernel speed up is small. We have observed that the speedup factors of the advection and collision kernels are the same for both grids of different size. This is an indication that the kernels are properly implemented and no degradation of performance is observed for grids of different size. This fact is particularly important for dynamically adapting grids with the cell count changing by up to 2 orders of magnitude during a simulation.

**TABLE 1.** UFS-LBM GPU speedup observed on Tesla C2070 and K20c cards.

| GPU Card | C2070 | K20c | C2070 | K20c |
|---|---|---|---|---|
| Spatial grid size (cells) | 100K | 100K | 1M | 1M |
| Advection speedup | 36 | 59 | 37 | 58 |
| Collision speedup | 150 | 151 | 158 | 155 |
| Ratio Advection/Collision | 2.3 | 2.4 | 2.3 | 2.4 |

We have not tried LBM solver on multi-GPU systems. However, we do not see any serious obstacles on the road. It is also expected that the gain for LBM-GPU should also be close to ideal, as for the Boltzmann-GPU, because of similar parallelization algorithms used in both modules.

Further acceleration of the LBM solver can be achieved using the regular-patch approach of GAMER.[24] The CPU parallelism could be based on rectangular domain decomposition with all patches within a rectangular sub-domain calculated by one CPU-GPU combination. Boundary conditions of each sub-domain could be provided by allocating the buffer patches surrounding the sub-domain. The GPU implementation, relies on 1) each mesh patch can be calculated independently as long as its ghost-zone values are provided (one thread block can be used to calculate one patch) and 2) all cells inside a patch can be calculated in parallel as long as there is a synchronization mechanism to coordinate the data update of each cell. This can be accomplished by using multiple CUDA threads to calculate different cells within the same patch, and store the updated results in the shared memory.

# 5. DIRECT SIMULATION MONTE CARLO

The DSMC module in UFS utilizes single mesh for particle transport, inter-particle collisions, statistics collection and visualization.[17] The AMR capabilities based on local flow properties (the particle mean free path, gradients of gas density, mean directed velocity and temperature, embedded surface curvature, etc.) can efficiently perform the same functions as transient subcells or other collision partner selection options available in modern DSMC codes.[47] The tree-based data structure of UFS-DSMC allows straightforward and efficient dynamic grid refinement and coarsening and removes the necessity of using subcells and additional models for collisional partners' selection. We demonstrate the benefits of UFS-DSMC for simulations of rarefied flows over moving bodies and describe algorithms for porting UFS-DSMC to GPUs.

## 5.1 DSMC with Octree Cartesian Mesh

The particle storage in UFS-DSMC is cell-based, which means that each cell has a list of particles assigned to it. This allows efficient selection of collisional partners for the collision step. During the free flight, when a particle crosses a face of a cell, it is removed from the list of its old cell and added to the list of a neighbor cell. The particles move inside each computational cell until they hit a cell face or a wall/boundary face. The particle tracing is efficient since all full cells are square or cubic; in cut cells, an additional solid face is checked for possible reflection and appropriate boundary conditions are applied when this face is crossed. Since a cell neighbor of a face which has been hit by a particle can be at a higher level, a new particle cell location is assigned depending on which part of the face has been traversed. During refinement of a coarse cell, particles are stored in new cells according to their position in this coarse cell. During cell coarsening, particles in fine cells are assigned to a new coarse cell. No special treatment (such as assigning different weights) is carried out to account for different volumes of cells on the adapted grid. Moreover, no special treatment of small cut cells is performed. The latter can lead to significant statistical scatter of the surface data (such as skin friction and heat flux). To tackle this issue, we have implemented special routines which traverse all cut cells at a given coarse level (usually corresponding to initial uniform grid). Inside each such (non-leaf) cell, all leaf cut cells are visited and the surface data are weight averaged with the corresponding cut-face surface area. This procedure allows significant reduction of statistical scatter in the calculated surface fluxes and provides similar advantages to a decoupling of surface elements from the grid structure as is usually done in other DSMC codes.

The UFS solvers deal with boundary cells using two techniques: 1) cut-cell technique combined with cell merging [14] and 2) Immersed Boundary Method (IBM).[48] UFS-DSMC uses the cut-cell technique. For moving-body capabilities, we utilize the corresponding module in the GFS framework which is based on the GTS library[49] to represent static or moving immersed solid objects. When a GTS-represented body is moved, vertices of each surface-mesh triangle are displaced by a small amount $\Delta \vec{r} = \vec{V}_{body} \Delta t$, where $\vec{V}_{body}$ is the surface velocity and $\Delta t$ the integration time step. The old surface grid at a previous time step $t - \Delta t$ and the new surface grid at the current time step $t$ are then stored, and the new volume grid is re-cut (to achieve second order accuracy in time, a surface grid from a time step $t - 2\Delta t$ is also stored). During this grid re-cutting step, the new volume grid undergoes the procedure called "hole cutting," in which one of the following 3 situations are realized: 1) position of a solid panel changes in cut cells, 2) new cut or full cells are created and 3) some cells become fully embedded into the solid and are thus destroyed. In the UFS-DSMC Module, particles (which are characterized by position, velocity, internal energy, etc.) are stored in each grid cell including the cut cells; such data storage allows straightforward implementation of the moving-body capabilities. Namely, when a solid panel moves in a cut cell, those particles which find themselves inside the solid are considered as having collided with the wall during the solid panel displacement and thus having acquired the wall velocity $\vec{V}_{body}$; the new particle position is then assumed to be close to the new wall position with the wall panel being slightly displaced away from it. When a new cell is created, it is assumed that no particles are initially present there. Once a new solid body position is set, particles are allowed to freely move and collide with other particles and with the (new and old) solid panels. When a particle collides with a solid panel (with a given temperature $T_{body}$), it is set to acquire the local translation velocity of the solid panel $\vec{V}_{body}$ (and its temperature $T_{body}$). During the next time step, the procedure repeats and so on.

## 5.2 UFS-DSMC simulations of rarefied flows over moving objects

Simulation of rarefied gas flows over moving objects is a challenging task important for several applications.[50,51] In this section, we show results of UFS-DSMC simulations of rarefied flows over moving objects to demonstrate benefits of the AMR technique. Figure 11 shows two examples of 2D simulations of particles of different shape moving through a monatomic gas (Argon) at rest with a density of $8.5 \times 10^{19}$ m$^{-3}$ and temperature of 200 K. The left part shows a star-shaped particle moving with a rotational speed of 2,000 rad/s and translational speed of 1,000 m/s. The right part shows two ellipsoids moving with a rotational speed of 2,000 and 1,500 rad/s, correspondingly, and the same translational speed of 1,000 m/s. These conditions correspond to Mach ~ 3, Kn ~0.1 flow. The grid adapts to dynamically resolve the shock structure and the Knudsen boundary layer around the bodies based on gradients of gas density and temperature. One can see in Figure 11 that the adapted grid allows efficiently resolving flow features by adding computational cells in the regions where it is necessary and coarsening the grid in the regions with small flow gradients to reduce the particle count. We have found that at least 10M particles are required to run this transient case to achieve acceptable statistics.

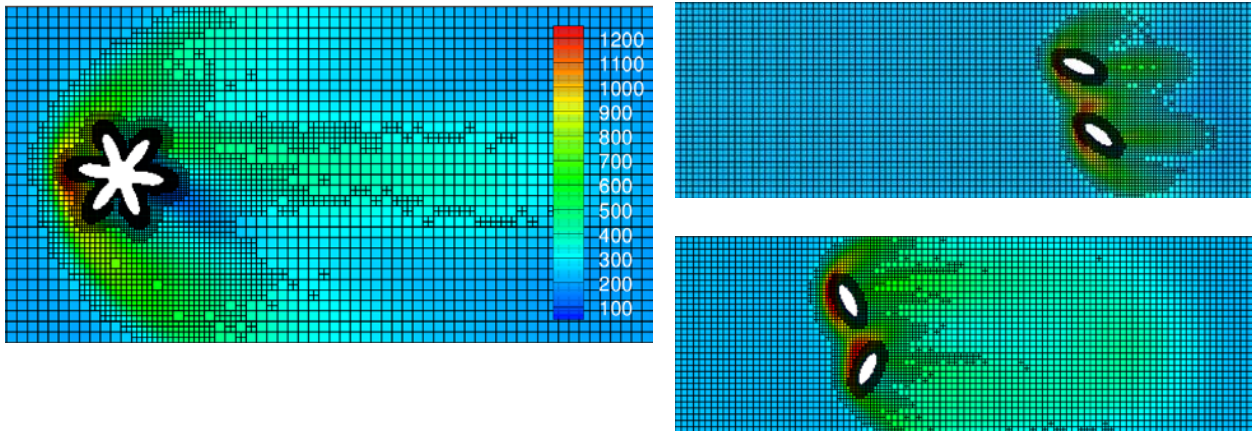

*Figure 11. Adapted computational grid colored by the translational temperature for moving star-shaped body (left) and two ellipsoids in relative motion (right).*

Parallel capabilities of UFS-DSMC were implemented using the FOT methodology described above. Figure 12 shows an example of parallel 3D simulations of an ellipsoid moving with a rotational speed of 2,000 rad/s and translational speed of 1,000 m/s through a gas at rest with the same properties as described above. This figure shows also an instantaneous PID distribution for the 8 CPU run. We have observed a close to ideal scaling with increasing number of processors because of close-to-uniform spatial distribution of the gas particles for this problem. For problems with highly non-uniform particle distributions in space, no such scaling is expected because no DLB is currently implemented for the UFS-DSMC-FOT. Implementing such a capability requires new graph partitioning and data structures to store/read in boxes during DLB.

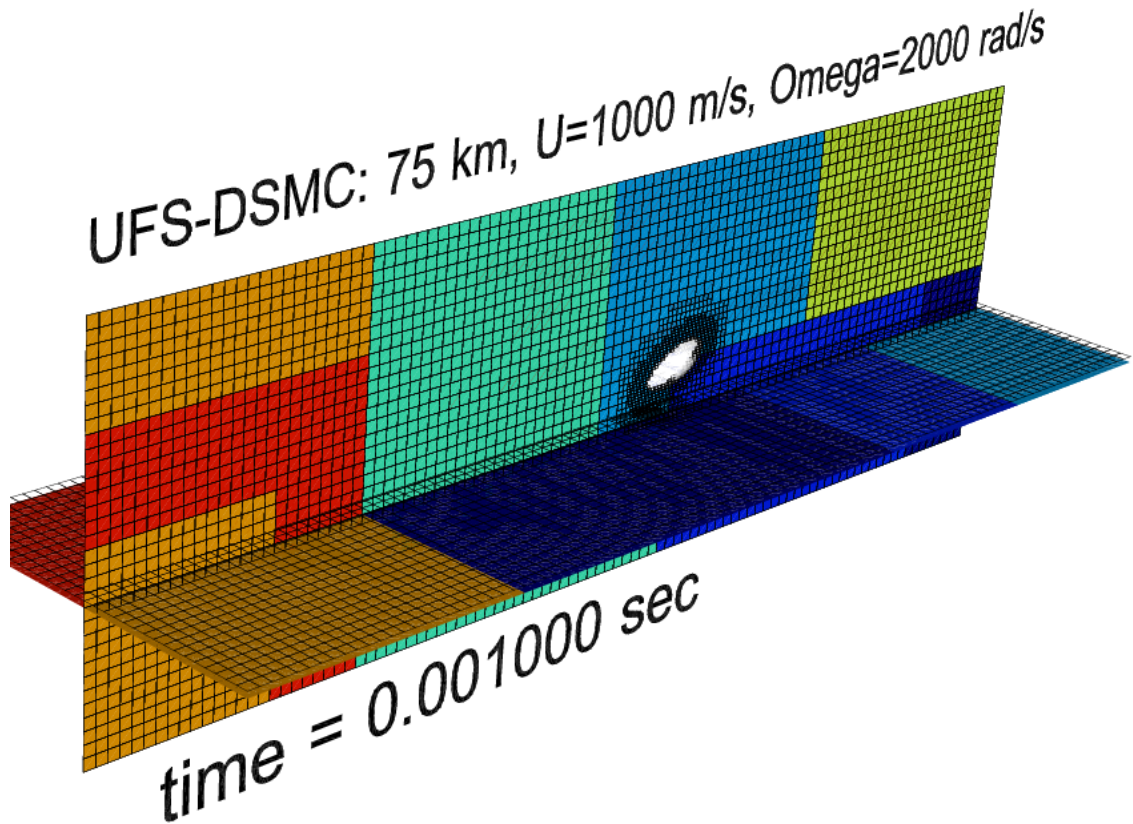

*Figure 12. Adapted computational grid colored by instantaneous Processor ID (PID) numbers for 8 CPU simulations of moving ellipsoid with 65M particles.*

### 5.3 Single-GPU Implementation of the DSMC Solver

A single-GPU implementation of UFS-DSMC was first reported in Ref. [29]. An all-device (i.e., GPU) computational approach [52] was adopted where the entire computation is performed on the GPU device, including particle moving, indexing, particle collisions and state sampling. Communication between the GPU and host (CPU) is only performed to enable multiple-GPU computation.

Currently, DSMC-GPU works only on static meshes. In other words, DSMC-GPU does not work with a) dynamic AMR and b) moving bodies. However, it is straightforward to unload the data from GPU to CPU every time the mesh changes (AMR takes place on CPU) and then load it back after properly resetting on new grid, as done in Boltzmann-GPU. Although this procedure might slow the GPU efficiency for the cases when the grids change frequently (as in the case for moving body), it should provide a simple solution for static non-uniform meshes.

### CUDA Kernels for Particle Movement, Collision and Sampling

Each part of DSMC-GPU was realized via a corresponding GPU kernel. Execution of a code on GPU requires that all data available on CPU host be prepared in a way that the GPU device can handle it. So, special data structures are required to be implemented. These data contain such information as simulation parameters, grid, cell, faces, neighbors, types of cells, faces, etc. On a host CPU, such information can be accessed via pointers and on GPU, it has to be accessed via indices.

Using the prepared cell and face data fields, particle movement is executed. Each particle is followed by a separate thread until the particle hits a face of a cell in which it is currently located. At cells faces, particles are reflected (solid faces or symmetry/diffusion reflection boundaries) and they are moved to its neighbor using neighbor indices (on coarse-fine interfaces, the new particle cell is decided depending on the location where the particle hits the face). The particle movement is carried out for old particles (already present in the domain) as well as for new particles generated at inflow/inlet boundaries.

When collisions are activated, collisions in each cell are treated by a separate thread. Particles are indexed and collision pairs are selected in each cell; new particle velocities are calculated based on the conservation laws. During particle movement and collisions, sampling of particle locations and velocities needs to be performed. For steady-state simulations, this is usually done starting from some start up step during a simulation. The particle sampling kernels are implemented using cell-based kernels (each cell is treated by a separate thread). Shared memory is used to efficiently calculate the required sums of particle quantities over all particles in each cell.

### 5.4 Multi-GPU Implementation of the DSMC Solver

For hybrid MPI-CUDA parallelization paradigm, domain decomposition was performed using the FOT method. In this method, the MPI protocol is used to exchange data from memory of all MPI processors and synchronize. CUDA is used to put the DSMC-related simulation components on GPU, and for data transfer between CPU (host) memory and GPU (device) global memory. For data exchanges between global memory of different GPU devices (for example, from device-A to device-B) we use the CUDA API function cudaMemcpy() to transfer data from device-A to host-A. Next, the data is transferred from host-A to host-B using the MPI protocols. Finally, we use cudaMemcpy() to transfer data from host-B to device-B. This flexible method allows a single subroutine to manage communications without concern for (i) the device type, or (ii) if the device is located on the same physical node (see Ref.[52] for more details).

The implementation of MPI-CUDA capabilities for DSMC followed several steps:
- A GPU kernel to calculate positions of all particles over a time step dt.
- Each particle's deterministic motion is handled by a CUDA thread, using data held entirely in global memory.
- Particles determined to have left the current GPU's simulation domain are placed into a buffer (in the device, and finally on the host) in preparation for migration to other GPU devices.
- Introduce new particles based on inlet boundary conditions.
- Send and receive from buffers of all MPI processors (host) with the MPI API protocols MPI_Send() and MPI_Recv().
- Reallocate particles from buffers into their newly allocated GPU's global memory.

Multi-GPU scaling was studied on the NASA Pleiades cluster with 64 Westmere nodes with one NVIDIA Tesla M2090 GPU per node. Each GPU with high speed memory is connected to the Westmere node via a PCI Express bus, and the nodes are connected via high-speed Infiniband. Figure 13 shows the real computational times (per 2,500 time steps) as a function of the number of GPUs. One can see that a very good scaling is obtained with the power law scaling factor being 0.8 (factor of 1 means ideal scaling). This proves a good efficiency of the hybrid MPI-CUDA algorithm for DSMC.

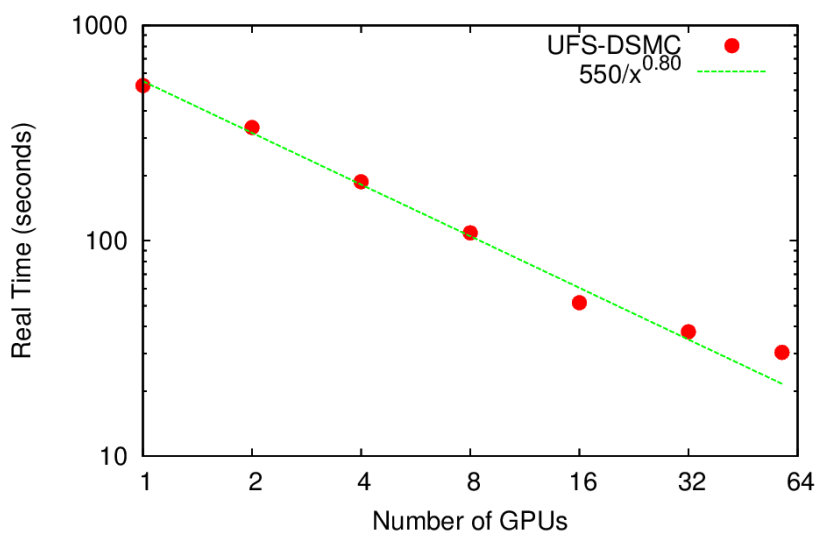

*Figure 13. Computing time as function of number of GPUs for the DSMC module in UFS.*

## 6. CONCLUSIONS AND OUTLOOK

Porting adaptive multi-scale kinetic-fluid solvers to heterogeneous CPU-GPU computing architectures poses several challenges associated with irregular data structures of AMR, vastly different costs of computing in fluid and kinetic cells, and dynamic balancing of CPU and GPU loads. We have described the simplest approach where GPUs were used for the kinetic cells only whereas the fluid cells were computed on CPUs. This methodology was demonstrated for UFS simulations of mixed rarefied-continuum flow over a hypersonic scramjet on a CPU-GPU cluster. We have achieved a computational breakthrough by using total number of adapted computational cells about 3.5M, and total number of Boltzmann cells about ~1.2M. Solving Boltzmann equation with a million of spatial cells appeared unrealistic a few years ago.

For the first time, we have described algorithms combining AMR and GPU capabilities for three kinetic modules in UFS: the DVM Boltzmann solver, the DSMC module, and the LBM solver. This implementation has taken advantages of the octree, dynamically adaptive grids with capabilities to automatically mesh solid objects of complex shape, moving-body capabilities, fine-grained (SFC) and coarse-grained (FOT) parallel capabilities with efficient dynamic load balancing among CPUs and GPUs. The DVM Boltzmann solver was parallelized using the SFC algorithm, the DSMC module was parallelized using the FOT algorithm. Double digit speedups on single GPU and good scaling for multi-GPU have been demonstrated.

For the first time, we developed an LBM solver with both AMR and GPU capabilities. Warp-based data treatment in GPU kernels for the advection and collisional modules have taken into account peculiarities of the LBM code on spacetree meshes. We have observed speedups of ~60x for the advection kernel and ~150x for the collisional kernel. The implementation demonstrated good scalability on grids of different size.

Future work will focus on resolving contradicting requirements of AMR and GPU. Improvement of the GPU-AMAR algorithm could be achieved by combining advantages of octree-type meshes with regular patches (blocks) desirable for GPU computing. We plan to implement the regular patch-based methodology of GAMER in the UFS framework. In each leaf computational cell of the irregular quad/octree grid, a regular (structured) patch will created of size 8×8 in 2D or 8×8×8 in 3D (other sizes are also possible). In order to be able to dynamically tailor the block size to the compute characteristics (e.g., GPU device type), we plan to implement a spacetree-block coupling by allowing different block sizes per spacetree node such as a 4×4×4 block used in NIRVANA[53] or even a general regular $N{\times}M{\times}L$ block, as it is implemented in the kinetic DVM solvers with velocity grid. This is expected to increase the GPU efficiency of the AMAR framework.

## ACKNOWLEDGMENTS

The work of Sergey Zabelok was supported by the Program No. 15 of the Presidium of Russian Academy of Sciences. The work at CFDRC was partially supported by the DoE SBIR Project DE-SC0010148 and by the DARPA SBIR Project W31P4Q-15-C-0047. We wish to thank Dr Martin Burtscher for useful discussions and suggestion of the warp algorithm for the LBM solver.